\documentclass[pre,twocolumn]{revtex4}

\usepackage{dcolumn}
\usepackage{amsmath}

\usepackage{graphicx}
\usepackage{amssymb}

\def\d{{\rm d}}
\def\O{{\rm O}}
\def\etal{{\it{}et~al.}}
\def\half{\mbox{$\frac12$}}
\def\eref#1{(\protect\ref{#1})}
\def\av#1{\langle#1\rangle}

\newlength{\figurewidth}
\begin{document}

\title{Convergence of threshold estimates for two-dimensional percolation}
\author{R. M. Ziff}
\affiliation{Michigan Center for Theoretical Physics and Department of
Chemical Engineering,\\
University of Michigan, Ann Arbor, MI 48109--2136}
\author{M. E. J. Newman}
\affiliation{Department of Physics, University of Michigan, Ann Arbor,
MI 48109--1120}
\affiliation{Santa Fe Institute, 1399 Hyde Park Road, Santa Fe,
NM 87501}

\begin{abstract}
  Using a recently introduced algorithm for simulating percolation in
  microcanonical (fixed-occupancy) samples, we study the convergence with
  increasing system size of a number of estimates for the percolation
  threshold for an open system with a square boundary, specifically
  for site percolation on a square lattice.  We show that the
  convergence of the so-called ``average-probability'' estimate is
  described by a non-trivial correction-to-scaling exponent as predicted
  previously, and measure the value of this exponent to be $0.90\pm0.02$.
  For the ``median'' and ``cell-to-cell'' estimates of the percolation
  threshold we verify that convergence does not depend on this exponent,
  having instead a slightly faster convergence with a trivial analytic
  leading exponent.
\end{abstract}


\maketitle

\section{Introduction}
\label{sec1}
Percolation~\cite{StaufferAharony92} is one of the most fundamental and
widely studied systems in statistical physics.  Theoretical studies of
percolation models and applications of percolation theory to physical
systems have spawned thousands of papers over the last few decades.  Even
so, there are some substantial gaps in our understanding of percolation.
For example, we have at present no exact value for the position $p_c$ of
the percolation threshold for site percolation on that simplest of
two-dimensional lattices, the square lattice.  And in three dimensions we
have almost no exact results whatsoever.  Because of this, numerical
methods have played an important role in the study of percolation.  In this
paper we consider a class of methods for estimating $p_c$ for site
percolation using finite-size scaling, and show how various estimates of
$p_c$ in this class scale with varying system size in two dimensions.

The methods studied here for measuring $p_c$ are widely used and are all
based upon consideration of the crossing probability function
$R_L(p)$~\cite{StaufferAharony92,ReynoldsStanleyKlein80,Kirkpatrick80}.
This function gives the probability that a connected path crosses the
system from one boundary segment to another, at site occupation
probability~$p$ and system size or length-scale~$L$.  Some examples of
these estimates are:
\begin{enumerate}
\item The renormalization-group (RG) fixed-point estimate $p_{\rm RG}(L) =
  p^*(L)$, where $p^*$ is the solution to the
  equation~\cite{ReynoldsStanleyKlein80,YoungStinchcombe75}
\begin{equation}
R_L(p) = p.
\label{RG}
\end{equation}
\item The average value of~$p$ at which crossing first
  occurs~\cite{StaufferAharony92,ReynoldsStanleyKlein80}:
\begin{eqnarray}
p_{\rm av}(L)  =  \av{p} &=& \int_0^1 p R_L'(p)\d p\nonumber\\
              &=& 1 - \int_0^1 R_L(p) \>\d p,
\label{pav}
\end{eqnarray}
where the last equality follows from integrating by parts.  The prime
indicates differentiation with respect to~$p$.
\item The estimate $p_{R_c}(L)$ corresponding to the point where $R_L(p)$
  equals its universal infinite-system value $R_c\equiv R_\infty(p_c)$
  (which is determined by the system shape and boundary
  conditions)~\cite{Ziff92}:
\begin{equation}
R_L(p) = R_c\>.
\label{pRc}
\end{equation}
For a square system, where $R_c = {1\over2}$, this estimate $p_{0.5}(L)$
corresponds to the median of the distribution $R'(p)$.  A related estimate
$p_{h+v}(L)$ for rectangular systems is the value of $p$ at which the
horizontal and vertical crossing probabilities sum to
unity~\cite{YonezawaSakamotoHori89,LanglandsPPS92,LanglandsPS94}:
\begin{equation}
R_L^{(h)}(p) + R_L^{(v)}(p) = 1.
\label{phv}
\end{equation}
This estimate is identical to $p_{R_c}(L) = p_{0.5}(L)$ when the boundary is a
perfect square.
\item The estimate $p_{\rm max}(L)$ which is the value of $p$ at which
  $R_L'(p)$ reaches a maximum (or equivalently, where $R_L(p)$ is at its
  inflection point)~\cite{ReynoldsStanleyKlein80}:
\begin{equation}
R_L''(p) = 0.
\label{pmax}
\end{equation}
\item The cell-to-cell RG estimate, which is the point where two systems of
  different size have the same value of~$R$~\cite{ReynoldsStanleyKlein80}.
  One possible choice for this estimate, $p^{(1)}_{\rm c-c}(L)$, is the
  value of $p$ at which
\begin{equation}
R_L(p) = R_{L-1}(p),
\label{pcc}
\end{equation}
while a second, $p^{(2)}_{\rm c-c}(L)$, is the point at which
\begin{equation}
R_L(p) = R_{L/2}(p).
\label{pcc2}
\end{equation}
\end{enumerate}

In order to use these estimates to determine the threshold precisely, we
need to know the manner in which they converge to~$p_c$ as $L\to\infty$.
While it is possible to simulate very large systems for which finite-size
effects may be quite small, the statistics for such simulations are still
relatively poor because of the small number of samples that can typically
be generated.  In most cases, better results can be derived by doing more
simulations on smaller systems, and this requires that the finite-size
behavior is characterized accurately.

It is usually assumed (based upon very general scaling arguments) that all
finite-system estimates of the percolation threshold converge to the bulk
value $p_c$ as
\begin{equation}
p_{\rm est}(L) - p_c \sim c L^{-1/\nu}
\label{RGconv}
\end{equation}
where $c$ is a system-dependent constant, and $\nu$ is the exponent
governing the correlation length~$\xi$, such that $\xi\sim|p-p_c|^{-\nu}$.
(For the two-dimensional systems we will be looking at in this
paper~$\nu=\frac43$.)  Well known exceptions to this behavior are a few
highly symmetric, self-dual systems, such as bond percolation on a square
lattice with a square boundary, and site percolation on a triangular
lattice with a rhomboidal boundary; in both these cases $R_L(p)$ is
perfectly symmetric about $p=\frac12$ for all~$L$ and all the estimates
above give $p_c=\frac12$ exactly.  For these systems, one could consider
the constant~$c$ above to be zero.

In Ref.~\onlinecite{Ziff92}, however, it was argued that for non-self-dual
systems with a square boundary, such as site percolation on a square
lattice (where because of the non-duality the estimates show finite-size
effects), the convergence of most of those estimates is faster than given
by Eq.~\eref{pcc}.  This is an observation of some practical significance,
since this particular system (site percolation on a square lattice with a
square external boundary) is one of the most commonly studied systems in
percolation.  Similar arguments also apply to other symmetric
two-dimensional crossing problems, such as a system with a rhomboidal
boundary, which is commonly used when simulating triangular and honeycomb
lattices.

The arguments of Ref.~\onlinecite{Ziff92} were based upon the hypothesis
that
\begin{equation}
R_L(p) \sim f_0(x) + L^{-1} f_1(x) + \ldots
\label{rz}
\end{equation}
for large~$L$, where $x = (p-p_c) L^{1/\nu}$, $f_0(x)$ represents the
universal part of~$R$, and $f_1(x)$ represents the first-order correction
to the scaling limit.  The choice of~$L^{-1}$ as the leading order of the
correction was based on numerical measurements of~$R$ at~$p_c$, and can be
derived from the assumption that the system is effectively slightly
rectangular in shape, because of the different types of boundary conditions
applied along the two principal axes~\cite{Ziff96}.  For small~$x$, it was
assumed that
\begin{eqnarray}
\label{ziff92series0}
f_0(x) &=& a_0 + a_1 x + a_3 x^3 + \ldots\\
\label{ziff92series1}
f_1(x) &=& b_0 + b_1 x + b_2 x^2 + \ldots
\end{eqnarray}
where $a_0=\frac12$ by the symmetry and self-duality of the square
boundary.  The same symmetry also implies that $f_0-\frac12$ is an odd
function in $x$ and hence that $a_n=0$ for all even $n>0$, as above.  (To
see that $f_0-\frac12$ is odd, note that for the perfectly dual system of
bond percolation on a square lattice, $R-\frac12$ is an odd function of $x$
for all~$L$, including $L=\infty$, and by universality systems with other
underlying lattices must behave the same way in the scaling limit.)  In
Ref.~\onlinecite{Ziff92}, no particular assumption was made about the
behavior of~$f_1(x)$, other than its analyticity about~$x=0$.

Note that the form of $f_0(x)$ in Eq.~\eref{ziff92series0} is not entirely
universal, because the independent variable $x$ should incorporate a metric
factor which depends upon the underlying
lattice~\cite{AharonyHovi94,HuLinChen95}.  For convenience, however, since
we are considering only the one system of site percolation on a square
lattice in this paper, we do not include that factor here.

Once the above assumptions, Eqs.~\eref{ziff92series0}
and~\eref{ziff92series1} are made, the convergence of the various estimates
of $p_c$ is straightforward to analyze, and one finds that while the RG
estimate does indeed converge according to Eq.~\eref{RGconv} (a result that
has been verified in many numerical studies), the rest of the estimates
above should converge according to the faster behavior
\begin{equation}
p_{\rm est}(L) - p_c \sim c L^{-1-1/\nu},
\label{rzest}
\end{equation}
where the constant $c$ varies from estimate to estimate.  Simulations
reported in Ref.~\onlinecite{Ziff92} for systems of size up to
$1024\times1024$ sites verified this convergence for the estimate $p_{0.5}$
to high accuracy.  The estimates $p_{\rm c-c}$, $p_{\rm max}$ and $p_{\rm
  av}$ were studied in Ref.~\onlinecite{Ziff92} using only exact
enumeration results for systems of sizes up to $7\times7$ which give
polynomials for~$R$ (see the appendix), and while the behavior of these
results was found to be roughly consistent with~\eref{rzest}, the
uncertainly due to higher-order corrections was large.

Following the publication of Ref.~\onlinecite{Ziff92}, Aharony and
Hovi~\cite{AharonyHovi94,HoviAharony96} argued that the irrelevant scaling
variables in the renormalization-group treatment of percolation imply a
slower leading-order convergence of~$R_L$ to its infinite-system value,
characterized by an exponent~$\omega$, whose value was deduced from the
Monte-Carlo work of Stauffer~\cite{Stauffer81} to be about~$\omega=0.85$.
(Note that Aharony and Hovi used $\theta_1$ to denote the exponent we
call~$\omega$.)  A variety of series expansion results from the early 1980s
were also analyzed to give values for this exponent ranging from~0.89 to
over~1~\cite{AdlerMoshePrivman82,AdlerMoshePrivman83}.

The argument given by Aharony and Hovi implies that the leading terms in
the expansion of $R_L(p)$ should in fact be
\begin{equation}
R_L(p) \sim f_0(x) + L^{-\omega} f_\omega(x) + L^{-1} f_1(x) + \ldots
\label{ha}
\end{equation}
where
\begin{equation}
f_\omega(x) = c_1 x + c_2 x^2 + \ldots
\label{ahseries}
\end{equation}
Hovi and Aharony argued that the constant term $c_0$ is zero for a square
system, because at $p_c$ there are no correction terms for the
square-lattice, bond-percolation system, and corrections due to irrelevant
variables should be universal.  They also argued that $f_1(x)$ should be
even, so that $b_1=0$ in~\eref{ziff92series1}, also by symmetry and
self-duality.  They discussed various consequences of these assumptions,
and presented numerical evidence that the term containing the exponent
$\omega$ is indeed the leading correction term, by showing that the
behavior of $R_L(p)$ for large~$x$ (that is, for $p \ne p_c$) was better
fit with such a term than without it.  However, the procedure they used did
not allow them to determine the value of~$\omega$ accurately, because of
the numerical difficulty of finding $R_L(p)$ precisely for all $p$.
Furthermore, they did not study the convergence of the other estimates
given above.

Recently~\cite{NewmanZiff00,NewmanZiff01}, the present authors have shown
that quantities like $R_L(p)$ can be studied efficiently for all $p$ by
first finding the crossing probability $R_{L,n}$ in a ``microcanonical''
system of exactly~$n$ occupied sites, and then convolving with a binomial
distribution to derive results for the corresponding ``canonical'' system
thus:
\begin{equation}
R_L(p) = \sum_{n=0}^N \left({N \atop n} \right) p^n (1-p)^{N-n} R_{L,n},
\label{convolve}
\end{equation}
where $N = L^2$ for site percolation on a square lattice.  The
microcanonical crossing probability is found using an efficient
cluster-joining algorithm employing data structures based on trees and a
fast method is employed for checking for percolation on the fly during the
progress of the calculation.  (While many of the ideas incorporated in this
method were put forward
previously~\cite{Hu92,GouldTobochnik96,deFreitasLucenaRoux99,ShchurVasilyev00,Tarjan75,JanLookmanStauffer83},
this was the first time that all of these components were combined in this
way, for the purpose of finding $R$ efficiently.)  In
Ref.~\onlinecite{NewmanZiff00} we studied the function corresponding to $R$
for the probability of a cluster wrapping around the boundary of a periodic
system on a torus.  In the present paper, we describe how that method can
be implemented for crossing of an open system, and we report results from
some large-scale simulations.  The results allow us to determine accurately
the behavior of all of the estimates above, and to test the theoretical
predictions that follow from Eq.~\eref{ha}.  As we will see, the appearance
of the ``irrelevant'' term in the scaling of some estimates is confirmed,
and a new, more precise value of~$\omega$ is found.

The outline of the paper is as follows.  In Section~\ref{analysis} we
derive the expected scaling behavior of the various estimates of $p_c$,
assuming the form~\eref{ha}.  In Section~\ref{methods} we describe our
numerical method, and in Section~\ref{results} we present the results of
our simulations.  In Section~\ref{concs} we give our conclusions.

\section{Convergence of estimates}
\label{analysis}
If we assume Eq.~\eref{ha} to be a correct description of the behavior of
$R_L(p)$, it is straightforward to deduce the resulting convergence of the
various estimates for~$p_c$.  In the following paragraphs we derive the
leading correction term for each estimate, or leading two terms when their
powers are close to each other.
\begin{widetext}
\begin{enumerate}
\item The RG fixed-point: The relevant terms of Eq.~\eref{RG} are
\begin{equation}
\half + a_1 x + a_3 x^3 + \ldots = p_c + x L^{-1/\nu},
\end{equation}
which implies
\begin{equation}
p_{\rm RG}(L) = p_c + \left[ {p_c - \frac12\over a_1}
 - {a_3(p_c - \frac12)^3 \over a_1^4} +\ldots \right] L^{-1/\nu}
 + \O(L^{-2/\nu}).
\end{equation}
The term in brackets is the value of $x$ that is the solution to $f_0(x) =
p_c$.
\item Average-probability estimate: Eq.\ (\ref{pav}) gives
\begin{equation}
p_{\rm av}(L) = 1 - L^{-1/\nu} \int_{x_0}^{x_1} \left[ f_0(x)
                + L^{-\omega} f_\omega(x) + L^{-1} f_1(x) \right] \,\d x,
\end{equation}
where $x_0=-p_c L^{1/\nu}$ and $x_1=(1-p_c) L^{1/\nu}$, which are the
values of $x$ at $p=0$ and~1 respectively.  Noting that, since $f_0(x)$ is
odd about $f_0(0)=\half$ and approaches 1 as $x \to \infty$, we have
\begin{equation}
L^{-1/\nu} \int_{x_0}^{x_1} f_0(x) \>\d x \sim L^{-1/\nu} x_1 = 1 - p_c,
\end{equation}
and we then get
\begin{equation}
p_{\rm av}(L) = p_c + L^{-\omega-1/\nu} \int_{-\infty}^\infty f_\omega(x) \>\d
x
                + L^{-1-1/\nu} \int_{-\infty}^\infty f_1(x) \>\d x,
\label{pavconv}
\end{equation}
where we have extended the limits of the integrals to $\pm\infty$.  This
result is also implied by Eq.~(40) of Ref.~\onlinecite{HoviAharony96},
for~$n=1$.  The order of the next correction depends upon the higher-order
corrections to~\eref{ha}.
\item Median-$p$ estimate:  Eq.~\eref{pRc} gives
\begin{equation}
\half + a_1 x  + b_0 L^{-1} + c_1 x L^{-\omega}\ldots = \half,
\end{equation}
which implies
\begin{equation}
p_{0.5}(L) = p_c - {b_0 \over a_1} L^{-1-1/\nu} + \O(L^{-1-\omega-1/\nu}).
\label{pRcconv}
\end{equation}
\item Maximum estimate: Eq.~\eref{pmax} gives
\begin{equation}
6 a_3 x + 2 b_2  L^{-1} + (2 c_2 + 6 c_3 x) L^{-\omega} + \ldots = 0,
\end{equation}
which implies
\begin{equation}
p_{\rm max}(L) = p_c - {c_2\over3a_3} L^{-\omega-1/\nu}
                 - {b_2\over3a_3} L^{-1-1/\nu} + \O(L^{-2\omega-1/\nu}).
\label{pmaxconv}
\end{equation}
\item Cell-to-cell estimate: Eq.~\eref{pcc} gives
\begin{eqnarray}
& & a_0 + a_1 (p-p_c)L^{1/\nu} + b_0 L^{-1} + c_1 (p-p_c) L^{1/\nu-\omega} +
\ldots\nonumber\\
& &\hspace{5mm} = a_0 + a_1 (p-p_c)(L-1)^{1/\nu} + b_0 (L-1)^{-1}
   + c_1 (p-p_c) (L-1)^{1/\nu-\omega} + \ldots
\label{ccc}
\end{eqnarray}
which implies
\begin{equation}
p^{(1)}_{\rm c-c}(L) = p_c  + {b_0\over a_1}\,\nu L^{-1-1/\nu} +
  \O(L^{-1-\omega-1/\nu}).
\label{pccconv}
\end{equation}
Likewise, for  $p^{(2)}_{\rm c-c}$ we have
\begin{equation}
p^{(2)}_{\rm c-c}(L) = p_c + {b_0\over a_1 (1 - 2^{-1/\nu})} L^{-1-1/\nu}
                       + \O(L^{-1-\omega-1/\nu}).
\label{pcc2conv}
\end{equation}
\end{enumerate}
\end{widetext}
Thus, the scaling of $p_{\rm RG}$, $p_{0.5}$, and $p_{\rm c-c}$ is
unaffected to leading order by the presence of the exponent~$\omega$.
However, $p_{\rm av}$ and $p_{\rm max}$ are affected by $\omega$ to leading
order, scaling as $L^{-\omega-1/\nu}$, slightly slower than predicted in
Ref.~\onlinecite{Ziff92} (order $L^{-1-1/\nu}$).  Note that $b_1$ does not
enter in any of these results, so the leading scaling does not change if
$b_1$ is equal to zero, as argued to be the case in
Ref.~\onlinecite{HoviAharony96}.

\section{Procedure}
\label{methods}
We have performed simulations to test the scaling hypotheses above using
the algorithm described in Refs.~\onlinecite{NewmanZiff00}
and~\onlinecite{NewmanZiff01}.  Briefly, sites are occupied one by one in
random order starting with an empty lattice.  Occupied sites form
contiguous clusters, each of which is identified uniquely by the site label
of a chosen single site within the cluster, which we call the ``root
site.''  Other (non-root) sites within a cluster possess pointers that
point either directly to the root site, or to other sites within the
cluster such that by following a succession of pointers one can get from
any site to the root.  A newly added site is considered to be a cluster of
size one, which is its own root site, and bonds are then added between it
and any adjacent occupied sites.  The clusters to which sites at either end
of such a bond belong are identified by following pointers from them to
their corresponding root sites, and if the root sites found are different
we conclude that two different clusters have been joined by the addition of
the bond.  We represent this by adding a pointer from the root site of one
of the clusters to the root site of the other.  Smaller clusters are always
made subclusters of larger ones, and all pointers followed are subsequently
changed to point directly to the root of their own cluster.  The net result
is an algorithm that can calculate $R_{L,n}$ (and many other observable
quantities) for all values of $n$ in average running time which is of the
order the area of the lattice, or $\O(L^2)$ for a square lattice.

In our previous calculations using this algorithm we measured the
probability of the existence of a cluster that wraps around the periodic
boundary conditions of a toroidal lattice.  In this paper we are interested
instead in the existence (or not) of a cluster that spans an open system
along one given direction.  There are (at least) three efficient methods
for detecting spanning of this kind, two of which are described in detail
in Ref.~\onlinecite{NewmanZiff01} and all of which we have used in the
present work.  In the first method, we use the same pointer-based trick
that we used in Ref.~\onlinecite{NewmanZiff00} to detect wrapping with
periodic boundary conditions, but start out with an $(L+1)\times (L+1)$
lattice in which one horizontal row of sites is fixed to be permanently
empty and one vertical one is fixed occupied.  Occurrence of a wrapping
cluster in such a system is then exactly equivalent to occurrence of a
spanning cluster in the horizontal direction in an open system with
dimensions $L\times L$.

In the second method, two complete rows of sites at the top and bottom of
an open $(L+2)\times (L+2)$ lattice are fixed permanently empty, and two
columns of $L$ sites on the left and the right sides of the lattice are
fixed occupied.  The two columns of occupied sites form two initial
clusters on the lattice.  By following pointers from one site in each of
these clusters it is then simple to determine whether the two clusters have
been amalgamated by other sites added between them: if they have the same
root site they have been amalgamated, otherwise they have not.  Performing
this check after the addition of each site to the lattice, we can detect
when a spanning cluster on the $L\times L$ open lattice first appears.  The
two methods have comparable running times and give compatible results.  The
second is somewhat simpler to implement.

In the third method, which was used for the majority of the simulations
here, we consider an $L \times L$ open lattice and keep track of the
minimum and maximum $x$- and $y$-coordinates for sites in each cluster,
updating their values as necessary when clusters are joined.  When $x_{\rm
  max}-x_{\rm min} = L-1$ for a cluster, we know that the cluster spans the
lattice in the horizontal direction, and similarly for vertical crossing.
This method allows one to check for both crossing events simultaneously,
and it is also efficient and easy to program.  A similar method was used in
Ref.~\onlinecite{Machtaetal96} for simulations of the Ising model.

Since in the present calculation we are only interested in the existence or
not of a system-spanning cluster, we can stop the simulation once a
spanning cluster is detected, as spanning must also occur for all higher
values of~$n$.  This produces about a 40\% saving in running time.  Each
simulation then produces just a single number, the value of~$n$ at which a
spanning cluster first appears (or two numbers if we check for spanning in
both the horizontal and vertical directions).  Making a histogram of these
values over many runs of the algorithm, we derive an estimate of the
probability $P_{L,n}$ that the system first percolates when the number of
occupied sites reaches~$n$.  This probability is related to the desired
function $R_{L,n}$ according to $P_{L,n} = R_{L,n}-R_{L,n-1}$, and hence
\begin{equation}
R_{L,n} = \sum_{n'=0}^n P_{L,n'}.
\label{RnPn}
\end{equation}
Once the $R_{L,n}$ is determined, $R_L(p)$, the corresponding function in
the ``canonical'' percolation ensemble, is calculated from
Eq.~\eref{convolve}, with the binomial coefficients for large $N$ being
calculated by iterative multiplication~\cite{NewmanZiff01}.  The estimates
$p_{\rm RG}$, $p_{0.5}$, and $p_{c-c}$ for the percolation threshold are
then evaluated directly according to Eqs.~\eref{pav}, \eref{pRc},
and~\eref{pcc}.  The remaining estimate, $p_{\rm av}$, could be found
directly by performing a numerical integral over $R_L(p)$, but a better
method is to use the following exact formula:
\begin{eqnarray}
p_{\rm av} &=& 1 - \int_0^1 R_L(p) \>\d p\nonumber\\
           &=& 1 - \sum_{n=0}^N \Biggl({N \atop n} \Biggr) R_{L,n}
               \int_0^1 p^n (1-p)^{N-n} \>\d p\nonumber\\
           &=& 1 - {1\over N+1} \sum_{n=0}^N R_{L,n}.
\end{eqnarray}
Using Eq.~\eref{RnPn} this can also be written directly in terms of
$P_{L,n}$ as
\begin{equation}
p_{\rm av} = 1 - {1\over N+1} \sum_{n=0}^N \sum_{n'=0}^n P_{L,n'}
= {1 \over N+1} \sum_{n=0}^N n P_{L,n},
\label{pavmicro}
\end{equation}
which means that the canonical average position of the percolation
threshold is $N/(N+1)$ times the microcanonical average $(1/N) \sum n
P_{L,n}$ and no convolution is necessary to find its value.  Higher moments
of the distribution $R'$ can be found in a similar fashion.  For the second
moment, for example, we have
\begin{eqnarray}
\av{p^2} &=& \int_0^1 p^2 R_L'(p) \>\d p\nonumber\\
         &=& 1 - {2\over(N+1)(N+2)}
             \sum_{n=0}^N (n + 1) R_{L,n}\nonumber\\
         &=& {1\over(N+1)(N+2)} \sum_{n=0}^N n(n+1) P_{L,n}.
\end{eqnarray}
To find where $R''_L(p) = 0$ for the estimate $p_{\rm max}$, we make use of
the following result:
\begin{eqnarray}
R_L''(p) &=& \sum_{n=0}^N \left({N \atop n} \right) p^n (1-p)^{N-n}
R_{L,n}\times\nonumber\\
& & \hspace{-10mm} \biggl[ {n(n-1)\over p^2} - {2n(N-n)\over p(1-p)} +
 {(N-n)(N-n+1)\over(1-p)^2}\biggr].\nonumber\\
\label{R''}
\end{eqnarray}
The above three results, along with Eq.~\eref{convolve}, demonstrate further
the advantage of calculating $R_L(p)$ through the microcanonical~$R_{L,n}$:
quantities like $R_L(p)$ and $R_L''(p)$ can be calculated exactly at all
$p$, while $p_{\rm av}$ can be found without introducing any error through
numerical integration.

\begin{table*}[t]
\begin{tabular}{r|l|lllllll}
$L$ & $N_S$ & $p_{\rm RG}$ & $p_{\rm av} = \langle p\rangle$ & $p_{0.5}$ & $p_{\rm c-c}$ & $p_{\rm max}$ & $\sqrt{\langle(\Delta p)^2\rangle}$ & $R_L(p_c)$ \\
\hline
$2$   & (exact)            & $0.61803399$ & $0.53333333$ & $0.54119610$ &                   & $0.57735027$ & $0.22110832$ & $0.5792507$ \\
$3$   & (exact)            & $0.61926013$ & $0.55238095$ & $0.55929632$ & $0.62073447^{(1)}$ & $0.58030237$ & $0.18137908$ & $0.5667036$ \\
$4$   & (exact)            & $0.61935542$ & $0.56400919$ & $0.56972413$ & $0.61958378^{(1)}$ & $0.58439952$ & $0.15483466$ & $0.5555884$ \\
$5$   & (exact)            & $0.61809529$ & $0.57114567$ & $0.57581007$ & $0.61350605^{(1)}$ & $0.58675948$ & $0.1358442$  & $0.5475384$ \\
$6$   & (exact)            & $0.61658709$ & $0.57585067$ & $0.57970276$ & $0.60920876^{(1)}$ & $0.58825653$ & $0.12151246$ & $0.5414670$ \\
$7$   & (exact)            & $0.61511736$ & $0.57911947$ & $0.58235130$ & $0.60607599^{(1)}$ & $0.58926561$ & $0.11027224$ & $0.5367513$ \\
$7$   & $10.0\times10^{10}$ & $0.6151180$  & $0.5791204$  & $0.5823519$  & $0.6060812^{(1)}$  & $0.5892655$  & $0.1102720$  & $0.536749$  \\
$8$   & $60.0\times10^9$    & $0.6137656$  & $0.5814866$  & $0.5842394$  & $0.608314^{(2)}$   & $0.5899755$  & $0.1011925$  & $0.532998$  \\
$16$  & $20.0\times10^9$    & $0.6069022$  & $0.5887819$  & $0.5898858$  & $0.598828^{(2)}$   & $0.5920104$  & $0.0633761$  & $0.518117$  \\
$32$  & $20.0\times10^9$    & $0.6016319$  & $0.5914246$  & $0.5918352$  & $0.594825^{(2)}$   & $0.5926026$  & $0.0387203$  & $0.509535$  \\
$64$  & $30.0\times10^9$    & $0.5981485$  & $0.5923179$  & $0.5924657$  & $0.593413^{(2)}$   & $0.5927391$  & $0.0233379$  & $0.504890$  \\
$128$ & $10.0\times10^9$    & $0.5959837$  & $0.5926087$  & $0.5926613$  & $0.592952^{(2)}$   & $0.5927577$  & $0.0139703$  & $0.502476$  \\
$256$ & $40.0\times10^8$    & $0.5946742$  & $0.5927013$  & $0.5927208$  & $0.592808^{(2)}$   & $0.5927536$  & $0.0083343$  & $0.50124$   \\
\end{tabular}
\caption{Various estimates of $p_c$ from the simulations ($L \ge 7$),
where $N_{S}$ is the number of samples, and from exact expressions
($L\le7$).  The cell-to-cell estimate is $p_{\rm c-c}^{(1)}$ for
superscript ${(1)}$ and  $p_{\rm c-c}^{(2)}$ for superscript ${(2)}$.
Errors in the numerical results are generally in the last digit quoted.
\label{table1} }
\end{table*}

In the more familiar binary search method for finding
$R_L(p)$~\cite{StaufferAharony92}, $p$~is increased or decreased to narrow
the bounds on one's estimate of position of the percolation point for a
given realization of the disorder.  This search process is stopped after
some number $m$ of iterations, typically about~15, giving to a resolution
of $2^{-m}$ on the estimate for~$p_c$.  Thus, $R_L(p)$ is evaluated at only
a finite set of points, and this adds some uncertainty to the calculation,
beyond the basic statistical error.  Given that our microcanonical method
is also much faster than binary search due to its efficient cluster-merging
and percolation-checking, there seems no reason to use other methods.

\section{Results of simulations}
\label{results}
Simulations were carried out for $L = 7$, 8, 16, 32, 64, 128,~256.  The
results are given in Table~\ref{table1}, along with exact results from
exhaustive enumeration of states for small systems with $L\le7$.  The
polynomials from which the exact results are derived are listed in the
appendix.  We also conducted some simulations at $L=7$ to compare numerical
and exact results, and the agreement was found to be perfect within the
statistical accuracy of the simulations.  The pseudo-random number
generator used for the simulations was the four-tap feedback generator
known as {\tt R9689} or {\tt gfsr4}~\cite{Ziff98}.

Error analysis for the simulation data indicates that the estimates
of~$p_c$ are accurate to about six figures, and the values of $R(p_c)$ are
accurate to four or five figures, as indicated the table.  We also
simulated $2.4\times10^7$ samples for a system of size $L=512$, but the
statistical accuracy of the results was insufficient to add anything to the
present analysis.

Consider the results for the estimate $p_{\rm max}$, whose convergence is
non-monotonic.  Its value starts below $p_c$ for small systems, then goes
above $p_c$ as lattice size passes though $L\simeq100$, and presumably
converges to $p_c$ from above as $L\to\infty$.  Indeed, according to
Eq.~\eref{pmaxconv}, $p_{\rm max}$ has two correction terms with
closely-spaced scaling exponents $-\omega-1/\nu\simeq-1.65$ (using the
value of $\omega$ from below) and $-1-1/\nu=-1.75$; it appears that these
terms contribute more or less equally in the range of system sizes that we
are considering.  The observed behavior is consistent with
Eq.~\eref{pmaxconv} if $b_2<0$, $c_2>0$, and the two are are roughly
comparable in magnitude.  (Note that $a_3<0$ because $f_0'(x)$ is at a
maximum at $x=0$).  It is not possible to fit the exponents of
Eq.~\eref{pmaxconv} reliably to these data.

The rest of the estimates are all monotonic, and lead to reasonably
straight lines when viewed on a logarithmic plot of $|p_{\rm est} - p_c|$
vs.~$L$, reflecting the leading power-law behavior.  To provide a more
sensitive representation of our data, we calculate successive slopes
between pairs of points for systems of size~$\half L$ and~$L$, giving
values $-\bigl[\log|p_{\rm est}(L)-p_c|-\log|p_{\rm est}(\half
L)-p_c|\bigr]\big/\log2$ for the leading exponent for the various
estimates~$p_{\rm est}(L)$.  For these calculations we used the value
$p_c=0.5927462$ given in Ref.~\onlinecite{NewmanZiff00}, which is
consistent with the data presented here, but of somewhat higher precision
than these data would yield.

\begin{figure}[t]
\resizebox{\figurewidth}{!}{\includegraphics{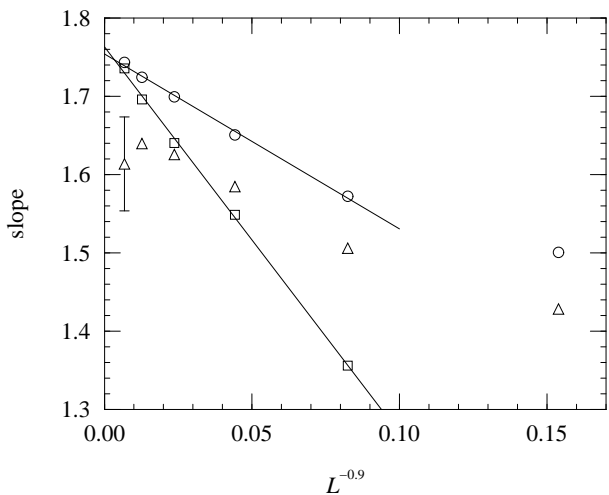}}
\caption{Values of the pairwise slopes $-\bigl[\log|p_{\rm est}(L)-p_c| - 
  \log|p_{\rm est}(\half L)-p_c|\bigr]\big/\log2$, plotted as a function
  $L^{-\omega}$, for the estimates $p_{0.5}(L)$ (circles), $p_{\rm av}(L)$
  (triangles), and $p^{(2)}_{\rm c-c}(L)$ (squares) with $p_c=0.5927462$.}
\label{fig1}
\end{figure}

In Fig.~\ref{fig1} we show the plots of the successive slopes for estimates
$p_{0.5}(L)$, $p^{(2)}_{\rm c-c}(L)$, and $p_{\rm av}(L)$, as a function of
$L^{-\omega}$ with $\omega=0.9$.  According to Eqs.~\eref{pRcconv}
and~\eref{pcc2conv}, both of these estimates should converge with a leading
exponent of $-1-1/\nu$ and a next-order term of order $-1-1/\nu-\omega$,
which implies that the successive slopes should fall on a straight line
when plotted as a function of $L^{-\omega}$, with intercept of~1.75.  This
behavior is indeed seen in Fig.~\ref{fig1}, with measured intercepts of
1.754 and 1.763 respectively.  (Note that agreement is not highly sensitive
to the value of~$\omega$; if the data were plotted as a function of
$L^{-1}$, the fit to linearity would not be much worse.)

The successive slopes for $p_{\rm av}(L)$ do not fall on as good a straight
line as the other estimates, presumably because the exponents
$-\omega-1/\nu$ and $-1-1/\nu$ of the two leading terms in the convergence
are closely spaced (see Eq.~\eref{pavconv}).  Extrapolation to $L=\infty$
is still possible however, and we find an intercept at
$1/\nu+\omega=1.65\pm0.02$, clearly different from the value of $1.75$ for
the other estimates, implying $\omega=0.90\pm0.02$, the figure we have used
above.  The error bars are smaller than the size of the symbols for all~$L$
except $L=256$, as shown in the plot.  Note that although $\omega$ is used
in the abscissa of Fig.~\ref{fig1}, its value has little effect on the
determination of $\omega$ from the intercept of~$p_{\rm av}(L)$.

\begin{figure}
\resizebox{\figurewidth}{!}{\includegraphics{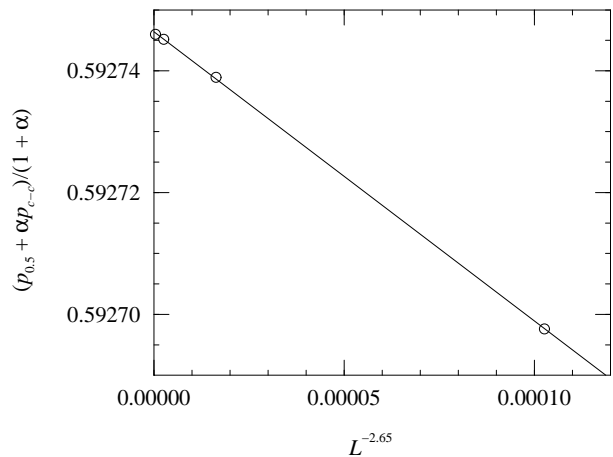}}
\caption{Plot of $(p_{0.5}+\alpha p^{(2)}_{\rm c-c})/(1+\alpha)$
  vs.\ $L^{-2.65}$ where $\alpha=1-2^{-1/\nu}$.}
\label{fig2}
\end{figure}

We can also compare the coefficients for the leading behavior of estimates
to their predicted values.  For example, the predicted value of the leading
coefficient for $p_{0.5}$, Eq.~\eref{pRcconv}, is $b_0/a_1=0.423$, using
values of $b_0=0.322$ (see below) and
$a_1=0.765$~\cite{Ziff92,HoviAharony96}.  This compares favorably with the
value measured here of~$0.436$.  And for $p^{(2)}_{\rm c-c}$, the
coefficient from~\eref{pcc2conv}, $b_0/[a_1(1-2^{-1/\nu})]$, should have a
value of $1.04$, which compares favorably with the measured value~$1.02$.
Note that if we take the linear combination of these two estimates (whose
finite-size corrections are opposite in sign), $(p_{0.5} + \alpha
p^{(2)}_{\rm c-c} )/(1+\alpha)$ where $\alpha = 1 - 2^{-1/\nu}$, the
leading correction terms are predicted to cancel one another and the
combination should have leading scaling of $L^{-1-\omega-1/\nu}=L^{-2.65}$.
And indeed this combination is seen to converge very quickly in our
numerical results, with values 0.592698, 0.592739, 0.592745, and 0.592746
for $L=32$, 64, 128, and~256 respectively.  The plot of these figures vs.\ 
$L^{-2.65}$ given in Fig.~\ref{fig2} shows linear behavior as expected,
with an intercept at $p=0.5927464(5)$, consistent with the best current
figure of $p_c=0.5927462(1)$~\cite{NewmanZiff00}.  Thus, by taking a
combination of estimates, we can improve the convergence rate for the open
system to the point where it becomes competitive with that of the periodic
system, in which the estimate with the best convergence has
exponent~$-\frac{11}{4}$~\cite{NewmanZiff00}.  This cancellation of leading
order corrections between the two terms is expected to be universal.

\begin{figure}
\resizebox{\figurewidth}{!}{\includegraphics{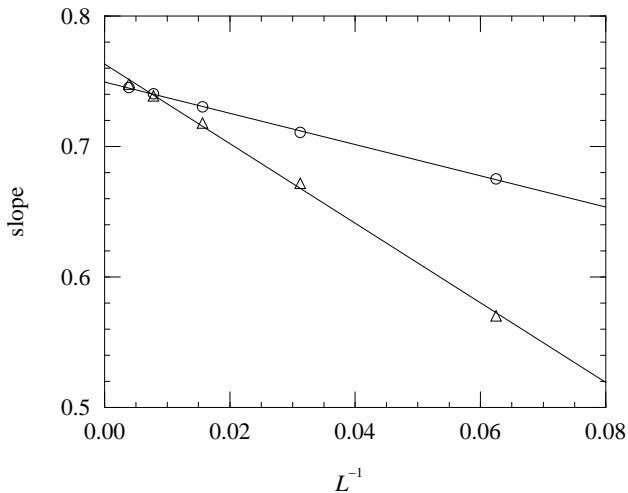}}
\caption{Values of the pairwise slopes of
  $-(\log|p_{\rm RG}(L)-p_c|-\log|p_{\rm RG}(L/2)-p_c|)/\log2$, with
  $p_c=0.5927462$ (triangles), and of the standard deviation of the
  distribution of~$p$ (circles), plotted as a function $L^{-1}$.  The lines
  are fit through all the points.}
\label{fig3}
\end{figure}

The results for $p_{\rm RG}$ and the standard deviation $\sigma =
\sqrt{\av{(\Delta p)^2}} = \sqrt{\av{p^2} - \av{p}^2}$ converge to the
infinite system value with exponent $1/\nu = -0.75$ as shown in Fig.\ 
\ref{fig3} where we plot the successive slopes as a function of $1/L$.
(The convergence of the standard deviation of the distribution is predicted
in Ref.~\onlinecite{HoviAharony96} and also follows from the equations of
Section~\ref{analysis}).  The intercepts are at 0.749 and 0.763
respectively.  We also show in Fig.~\ref{fig3} a plot of the type
introduced by Stauffer, in which $p_{\rm RG}$ is shown as a function
of~$\sigma$, allowing extrapolation to infinite system size to be carried
out without knowledge of the value of~$\nu$~\cite{StaufferAharony92}.  This
plot shows nearly linear behavior, with the last three points ($L=64$, 128,
and 256) well fit by the line $p_{\rm RG} = 0.5927465 + 0.231512\sigma$
with $R^2=0.9999979$.  The intercept is in excellent agreement with the
known value of~$p_c$, although this agreement is perhaps fortuitous,
considering the slow convergence of the RG estimate.

The last column of Table~\ref{table1} gives the crossing probability
at~$p_c$.  The considerations in Section~\ref{analysis} imply that
$R_L(p_c) \sim 1/2 + b_0/L$ with no contributions from the irrelevant
scaling variable~\cite{Ziff92,Ziff94,HoviAharony96}, and indeed an analysis
of this data shows good agreement with the behavior $R_L(p_c) = 0.320/L -
0.44/L^2 +\ldots $, yielding $b_0 = 0.320\pm0.001$.  This is nearly
identical to the value $0.319$ given in Ref.~\onlinecite{Ziff92} (where
larger systems, but with lower statistics, were generated) and the value
$0.31\pm0.01$ of Ref.~\onlinecite{HoviAharony96}.

\section{Conclusions}
\label{concs}
We have studied the finite-size scaling of estimates of the percolation
threshold $p_c$ derived from the crossing probability $R_L(p)$ for site
percolation on the square lattice.  Our numerical results confirm that
different estimates converge to $p_c$ with a variety of scaling exponents
as predicted by the scaling theory developed in Refs.~\onlinecite{Ziff92},
\onlinecite{AharonyHovi94}, and~\onlinecite{HoviAharony96}.  In particular,
we have shown that the average threshold estimate $p_{\rm av}$ converges
with a nontrivial exponent $L^{-\omega-1/\nu}$ whose origins lie in the
irrelevant variables in the renormalization group treatment of the problem,
and our results for this estimate provide us with a direct measurement of
that exponent.  We find $\omega=0.90\pm02$, somewhat higher than the value
of about 0.85 found previously~\cite{Stauffer81} but consistent with the
(wide) bounds set by series
studies~\cite{AdlerMoshePrivman82,AdlerMoshePrivman83}.  Our result is in
good agreement with a renormalization-group result of $\omega=0.915$ found
by Burkhardt~\cite{Burkhardt80,note1}.  More extensive simulations could be
done to give~$\omega$ to higher precision.

Of the estimates considered here, $p_{\rm av}$ is the only one that shows
the effect of the irrelevant exponent clearly.  The estimates $p_{0.5}$ and
$p_{\rm c-c}$ are confirmed to converge as $L^{-1-1/\nu}$, as proposed
previously~\cite{Ziff92}.  The maximum estimate $p_{\rm max}$ is found to
exhibit non-monotonic behavior, which can be explained by competition
between correction terms with closely spaced exponents $-1-1/\nu$ and
$-\omega-1/\nu$.

The numerical results reported here were found using a microcanonical
simulation method which allows one to calculate $R_L(p)$ easily for
any~$p$~\cite{NewmanZiff00,NewmanZiff01}.  The various estimates can then
be found quickly to any desired degree of precision by applying appropriate
formulas, Eqs.~(\ref{RG}--\ref{pcc2}).  This method proves particularly
advantageous for the estimate~$p_{\rm av}$, since this estimate depends on
knowing $R_L(p)$ for all values of~$p$, the determination of which by most
other methods requires a great deal of work.  From the microcanonical data,
$p_{\rm av}$ can be found without any calculational bias using
Eq.~\eref{pavmicro}.

Having characterized the convergence rates of our various threshold
estimates, one can go back to older literature and find many instances
where an anomalous rate of convergence was seen but not fully understood or
recognized.  For example, Reynolds~\etal~\cite{ReynoldsStanleyKlein80}
derived an estimate of $p_c$ which is equivalent to our $p_{\rm av}$ from
numerical data for the one-way crossing probability which they
denoted~$R_1$.  Assuming this estimate to scale as Eq.~\eref{RGconv}, they
plotted their results against $L^{-1/\nu}$ (their Fig.~13).  The resulting
plot is seen to fall on a nearly vertical line, consistent with
higher-order behavior.  Fitting their data with the supposed $L^{-1/\nu}$
scaling, they deduced a best estimate of $p_c=0.5931$ in the large
system-size limit.  If however one assumes instead the $L^{-\omega-1/\nu}$
scaling derived here, the intercept of their data becomes
$p_c\simeq0.5927$, which is much closer to the current best estimate of
this quantity.

\begin{figure}
\resizebox{\figurewidth}{!}{\includegraphics{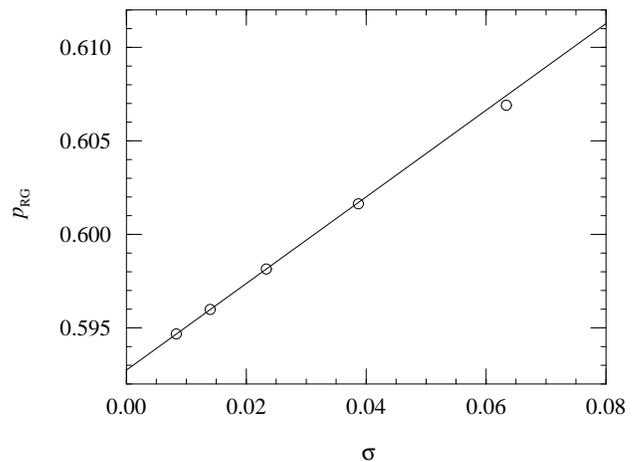}}
\caption{``Stauffer plot'' of $p_{\rm RG}$ vs.\ $\sigma =
  \sqrt{\av{(\Delta p)^2}}$; the line is fit through the leftmost three
  points and its equation is given in the text.}
\label{fig4}
\end{figure}

Yonezawa~\etal~\cite{YonezawaSakamotoHori89} plotted a quantity essentially
equivalent to our~$p_{0.5}$ as a function of $L^{-1/\nu}$, and found
apparent agreement with this assumed $L$-dependence (their Figs.~7 and~8).
The expected $L^{-1-1/\nu}$ behavior is evidently too weak to be
distinguished from $L^{-1/\nu}$ within the errors on their numerical data.
Similarly, Hu~\etal~\cite{HuChenWu96} believed the cell-to-cell estimate
$p_{\rm c-c}^{(1)}$ to be insensitive to~$L$; again, the precision of their
work did not allow them to observe the higher-order scaling predicted by
Eq.~\eref{pcc2conv}.

\begin{table*}
\begin{tabular}{l|l}
$L$ & $R_L(p)$ \\
\hline
2 & $2p^2 - p^4$ \\
3 & $3p^3 + 4p^4 - 6p^5 - 9p^6 + 14p^7 - 6p^8 + p^9$ \\
4 & $4p^4 + 12p^5 - 6p^6 - 28p^7 - 22p^8 + 48p^9 + 66p^{10} - 108p^{11} +
10p^{12} + 44p^{13} - 20p^{14} + p^{16}$ \\
5 & $5p^5 + 24p^6 + 12p^7 - 62p^8 - 92p^9 - 41p^{10} + 274p^{11} + 42p^{12}
+ 474p^{13} - 1336p^{14} + 172p^{15} - 197p^{16} + 4791p^{17} - 9015p^{18}$ \\
 & \hspace{3.65mm} ${} + 8013p^{19} - 4261p^{20} + 1559p^{21} - 450p^{22} + 103p^{23} - 15p^{24} + p^{25}$ \\
6 & $6p^6 + 40p^7 + 60p^8 - 80p^9 - 248p^{10} - 276p^{11} + 201p^{12} +
944p^{13} - 298p^{14} + 2392p^{15} - 2420p^{16} + 548p^{17} - 24848p^{18}$ \\
 & \hspace{3.65mm} ${} + 38688p^{19} - 7540p^{20} + 51564p^{21} - 117312p^{22} - 133312p^{23} + 588639p^{24} - 608464p^{25} + 68362p^{26} + 420396p^{27}$ \\
 & \hspace{3.65mm} ${} - 455910p^{28} + 235816p^{29} - 62454p^{30} + 3200p^{31} + 3212p^{32} - 1024p^{33} +
120p^{34} - p^{36}$ \\
7 & $7p^7 + 60p^8 + 150p^9 - 18p^{10} - 490p^{11} - 885p^{12} - 318p^{13} +
1464p^{14} + 3056p^{15} - 1586p^{16} + 5584p^{17} - 6520p^{18} + 43150p^{19}$ \\
 & \hspace{3.65mm} ${} - 153589p^{20} + 128504p^{21} - 407257p^{22} + 1278288p^{23} -
1243193p^{24} + 2195374p^{25} - 6983630p^{26} + 8271536p^{27}$ \\
 & \hspace{3.65mm} ${} - 6990669p^{28} + 17741331p^{29} - 11344431p^{30} - 55294929p^{31} +
91642905p^{32} + 67152194p^{33} - 374255572p^{34}$ \\
 & \hspace{3.65mm} ${} + 557174473p^{35} - 463108229p^{36} + 225338948p^{37} - 47135360p^{38} - 12950691p^{39} + 11168848p^{40} - 1724804p^{41}$ \\
 & \hspace{3.65mm} ${} - 1067305p^{42} + 689318p^{43} -
196565p^{44} + 34848p^{45} - 4391p^{46} + 422p^{47} - 28p^{48} + p^{49}$ \\
\end{tabular}
\caption{Exact results for $R_L(p)$ expressed as polynomials in~$p$, for
  $L=2\ldots7$.}
\label{rlptable}
\end{table*}

It should be emphasized that the convergence behavior discussed here is
specific to a two-dimensional system with a square or rhomboidal open
boundary, with crossing defined as a path from one specified side to its
opposite.  Because of the symmetry of this system, some terms cancel out,
allowing various higher-order corrections to become dominant.  For
different boundaries (such as rectangular ones), and for higher dimensional
systems, the behavior will generally be different.  In those cases, most of
our estimates will scale as the conventional $L^{-1/\nu}$, except perhaps
the estimates $p_{\rm c-c}$ and $p_{R_c}$.  (To employ the latter, $R_c$
must be known, but we have exact values only for rectangular and
conformally related two-dimensional systems~\cite{Cardy92}.)  Study of
these other systems is a subject for future research.

Another approach to measuring~$p_c$ is to use periodic rather than open
boundary conditions.  A partially periodic system in two dimensions is a
cylinder, and crossing in this system was studied in
Ref.~\onlinecite{HoviAharony96}.  The fully periodic rectangle is a torus,
and the criterion of crossing is replaced by criteria involving the
different topologically distinct ways in which clusters can wrap around the
boundaries~\cite{Pinson94}.  (Some
authors~\cite{HuLinChen95,Hu96,Machtaetal96} have also considered the
percolation criterion in which a cluster has the full dimension of the
lattice along at least one axis but does not necessarily wrap around.)  In
Ref.~\onlinecite{NewmanZiff01} we showed that many estimates of $p_c$ on
the torus converge a factor $L$ faster than estimates for the open
square---some converging as fast as~$L^{-2-1/\nu}$.

In conclusion, it is clear that the convergence of estimates for the
critical occupation probability $p_c$ in percolation systems is highly
dependent upon the nature of the estimate, as well as the shape and
boundary conditions of the system, and that the shrewd use of this fact can
allow one to make very accurate estimates of~$p_c$ and scaling exponents.

\appendix
\section{Exact enumeration results}
In Table~\ref{rlptable} we give the exact expressions for the crossing
probability function $R_L(p)$ for site percolation on a square lattice of
size $L\times L$, for crossing from one given side of the square to the
opposite side (such as left to right).  The results for $L=2$ to~5 were
given previously by Reynolds~\etal~\cite{ReynoldsStanleyKlein80}.  Those
for $L=6$ and $L=7$ were calculated previously for the work reported in
Ref.~\onlinecite{Ziff92}, but reported in a different format (results were
given for $R_L'(p)$ rather than $R_L(p)$ itself).  From the results here,
one can with reasonable ease calculate the various estimates of $p_c$ given
in Table~\ref{table1} using a symbolic manipulation program such as {\it
  Maple\/} or {\it Mathematica.}  A file containing these polynomials in
forms readable by such programs is available by email from the authors.

\begin{table*}
\begin{tabular}{l|l}
$L$ & $R_L(p,q)$ \\
\hline
2 & $2p^{2}q^{2} + 4p^{3}q + p^{4}$ \\
3 & $3p^{3}q^{6}+22p^{4}q^{5} + 59p^{5}q^{4} + 67p^{6}q^{3} +
36p^{7}q^{2} + 9p^{8}q + p^{9}$ \\
4 & $4p^{4}q^{12} + 60p^{5}q^{11} + 390p^{6}q^{10} + 1452p^{7}q^{9} +
3416p^{8}q^{8} + 5272p^{9}q^{7} + 5414p^{10}q^{6} + 3736p^{11}q^{5} + 1752p^{12}q^{4} + 560p^{13}q^{3}$ \\
 & \hspace{8.0mm} ${} + 120p^{14}q^{2}
 + 16p^{15}q + p^{16}$ \\
5 & $5p^{5}q^{20} + 124p^{6}q^{19} + 1418p^{7}q^{18} + 9958p^{8}q^{17}
+ 48171p^{9}q^{16} + 170391p^{10}q^{15} + 456051p^{11}q^{14} + 942077p^{12}q^{13}$ \\
 & \hspace{8.0mm} ${} + 1518133p^{13}q^{12} + 1917887p^{14}q^{11} +
1903359p^{15}q^{10} + 1486308p^{16}q^{9} + 915643p^{17}q^{8} +
446538p^{18}q^{7} + 172749p^{19}q^{6}$ \\
 & \hspace{8.0mm} ${} + 52871p^{20}q^{5} +
12650p^{21}q^{4} + 2300p^{22}q^{3} + 300p^{23}q^{2} + 25p^{24}q + p^{25}$ \\
6 & $6p^{6}q^{30} + 220p^{7}q^{29} + 3830p^{8}q^{28} +
42200p^{9}q^{27} + 330862p^{10}q^{26} + 1966832p^{11}q^{25} +
9220051p^{12}q^{24} + 34986568p^{13}q^{23}$ \\
 & \hspace{8.0mm} ${} + 109429240p^{14}q^{22} +
285726952p^{15}q^{21} + 628339894p^{16}q^{20} + 1170656172p^{17}q^{19}
+ 1854519856p^{18}q^{18}$ \\
 & \hspace{8.0mm} ${} + 2502797192p^{19}q^{17} +
2879547507p^{20}q^{16} + 2824773868p^{21}q^{15} +
2362953818p^{22}q^{14} + 1686455720p^{23}q^{13}$ \\
 & \hspace{8.0mm} ${} + 1028085197p^{24}q^{12} + 536110144p^{25}q^{11} + 239427498p^{26}q^{10}
+ 91584720p^{27}q^{9} + 29943238p^{28}q^{8} + 8322620p^{29}q^{7}$ \\
 & \hspace{8.0mm} ${} + 1946842p^{30}q^{6} + 376992p^{31}q^{5} + 58905p^{32}q^{4} +
7140p^{33}q^{3} + 630p^{34}q^{2} + 36p^{35}q + p^{36}$ \\
7 & $7p^{7}q^{42} + 354p^{8}q^{41} + 8637p^{9}q^{40} +
135542p^{10}q^{39} + 1538918p^{11}q^{38} + 13480033p^{12}q^{37} +
94850847p^{13}q^{36} + 551119224p^{14}q^{35}$ \\
 & \hspace{8.0mm} ${} + 2697329225p^{15}q^{34}
+ 11286245629p^{16}q^{33} + 40833575812p^{17}q^{32} +
128871332816p^{18}q^{31} + 357226485246p^{19}q^{30}$ \\
 & \hspace{8.0mm} ${} + 874366412699p^{20}q^{29} + 1897489913029p^{21}q^{28} +
3662042878777p^{22}q^{27} + 6298869803283p^{23}q^{26}$ \\
 & \hspace{8.0mm} ${} + 9669568447297p^{24}q^{25} + 13258506844289p^{25}q^{24} +
16242412033336p^{26}q^{23} + 17776880198790p^{27}q^{22}$ \\
 & \hspace{8.0mm} ${} + 17378859362974p^{28}q^{21} + 15172837588687p^{29}q^{20} +
11830013256560p^{30}q^{19} + 8239207757621p^{31}q^{18}$ \\
 & \hspace{8.0mm} ${} + 5128578282954p^{32}q^{17} + 2855162977558p^{33}q^{16} +
1422652678272p^{34}q^{15} + 634745588151p^{35}q^{14}$ \\
 & \hspace{8.0mm} ${} + 253562760568p^{36}q^{13} + 90598044853p^{37}q^{12} +
28888611591p^{38}q^{11} + 8189388138p^{39}q^{10} + 2052078152p^{40}q^{9}$ \\
 & \hspace{8.0mm} ${} + 450849373p^{41}q^{8} + 85897197p^{42}q^{7}
 + 13983816p^{43}q^{6} + 1906884p^{44}q^{5} + 211876p^{45}q^{4} +
18424p^{46}q^{3} + 1176p^{47}q^{2}$ \\
 & \hspace{8.0mm} ${} + 49p^{48}q + p^{49}$
\end{tabular}
\caption{Exact results for $R_L(p,q)$ expressed as polynomials in~$p$ and
$q\equiv1-p$, for $L=2\ldots7$.}
\label{rlpqtable}
\end{table*}

An alternative way to represent these results is as a series in $p^n
q^{N-n}$ where $q=1-p$ and $N=L^2$.  The transformation can be achieved by
substituting $p\to1/(1+r)$, multiplying by $(1+r)^N p^N$, expanding, and
replacing $r\to q/p$ again.  The results are given in
Table~\ref{rlpqtable}.

This is also the form that Reynolds~\etal~\cite{ReynoldsStanleyKlein80}
used in their series for $R_2$ through~$R_5$.  From the present point of
view, these series are interesting because they are in precisely the form
of Eq.~\eref{convolve}, so that the coefficient $c_{L,n}$ of $p^n q^{N-n}$
above is related to the microcanonical crossing probability $R_n$ simply by
\begin{equation}
R_{L,n} = {c_{L,n} \over {N \choose n}} = {c_{L,n}\,n! (N-n)! \over N!}.
\end{equation}
That is, $c_{L,n}$ represents the number of configurations with $n$
occupied sites that satisfy the crossing criterion, out of a total of
${N\choose n}$ possible configurations of the $n$ occupied sites among the
$N=L^2$ sites of the lattice.  For example, of the $16!/(6!10!)=8008$
possible configurations of 6 occupied sites on a $4\times4$ lattice,
exactly 390 are percolating by crossing in one direction (from the third
term in $R_4(p,q)$), yielding a microcanonical probability $R_{4,6} =
390/8008 = 0.048701\ldots$\ \ Likewise, for $n=16$ occupied sites on the
$4\times4$ system, there is exactly one percolating system out of one total
system.

Thus, the polynomials given in Table~\ref{rlpqtable} represent the
microcanonical $R_{L,n}$ for $L$ up to~7.


\begin{thebibliography}{99}
\bibitem{note1} In fact, Hovi and Aharony found that the data for $p_{\rm
    av}$ in their Fig.~1 could be fit best using a value of
  $\omega\simeq0.90$, but for the data of their Fig.~5 $\omega = 0.85$ gave
  the better fit.
\end{thebibliography}
\end{document}